\documentclass[preprint]{aastex}
\usepackage{natbib}

\shorttitle{GRAVITY}
\shortauthors{Blind et al.}

\begin{document}
\title{GRAVITY: the VLTI 4-beam combiner for narrow-angle astrometry and interferometric imaging}

\author{N.\ Blind\footnote{These authors made significant contribution regarding the work on fibers reported in this paper.} ,  F. Eisenhauer$^\star$, S. Gillessen$^\star$, Y. Kok$^\star$, M. Lippa$^\star$, L. Burtscher, O. Hans, M. Haug, F. Haussmann, S. Huber, A. Janssen, S. Kellner, T. Ott, O. Pfuhl, E. Sturm, J. Weber, E. Wieprecht}
\affil{Max-Planck-Institut f\"ur extraterrestrische Physik (MPE), 85748 Garching, Germany}

\author{G. Perrin$^\star$, R. Dembet$^\star$, P. F\'edou$^\star$, S. Lacour$^\star$}
\affil{LESIA, Observ. de Paris Meudon, 5, place Jules Janssen, 92195 Meudon Cedex, France }

\author{K. Perraut$^\star$, L. Jocou$^\star$}
\affil{Univ. Grenoble Alpes, IPAG, F-38000 Grenoble, France\\
CNRS, IPAG, F-38000 Grenoble, France}

\author{A. Amorim}
\affil{ SIM, Fac. de Ci\^encias da Univ. de Lisboa, Campo Grande, Edif. C1, P-1749-016 Lisbon, Portugal}

\author{W. Brandner}
\affil{Max-Planck-Institut f\"ur Astronomie, K\"onigstuhl 17, 69117 Heidelberg, Germany}

\author{C. Straubmeier}
\affil{ I. Physikalisches Institut, Universit\"at zu K\"oln, Z\"ulpicher Strasse 77, 50937 K\"oln, Germany}

\begin{abstract}
GRAVITY is the second generation Very Large Telescope Interferometer instrument for precision narrow-angle astrometry and interferometric imaging in the Near Infra-Red (NIR). It shall provide precision astrometry of order 10 microarcseconds, and imaging capability at a few milliarcsecond resolution, and hence will revolutionise dynamical measurements of celestial objects. GRAVITY is currently in the last stages of its integration and tests in Garching at MPE, and will be delivered to the VLT Interferometer (VLTI) in 2015. We present here the instrument, with a particular focus on the components making use of fibres: integrated optics beam combiners, polarisation rotators, fibre differential delay lines, and the metrology.
\end{abstract}


\section{Introduction}

We present in this paper GRAVITY, the second generation Very Large Telescope Interferometer instrument for precision narrow-angle astrometry and interferometric imaging. It shall provide precision astrometry of order 10 micro-arcseconds, as well as spectro-imaging capability at a few milli-arcsecond resolution. In this respect, it will revolutionise dynamical measurements of celestial objects. Here is a summary of the science cases of GRAVITY:
\begin{itemize}
\item {\bf Stellar orbits and flares around Sgr A*} -- The primary goal of the GRAVITY astrometric mode is to probe dynamical motions around the Galactic Center Super Massive Black Hole (SMBH). It will be able to measure deviation from Newtonian orbits caused by, e.g.: {\it 1)} A possible cluster of dark objects (e.g. neutron stars or stellar mass black holes) around Sgr A*, breaking the single point mass hypothesis; {\it 2)} Relativistic effects, like relativistic precession that could be seen in a few years observation time. If the current interpretation of the NIR flares from Sgr A* as localized events in the innermost region of the accretion flow is correct, GRAVITY has the potential of directly determining the space-time metric around this black hole, and to test General Relativity in the currently unexplored strong curvature limit.
\item {\bf Black holes} -- In the same vain, GRAVITY aims at detecting and measuring the mass of black holes in globular clusters throughout the Milky Way by means of stellar dynamics.
\item {\bf Active Galactic Nuclei (AGN)} -- When the sphere of influence of SMBH in AGNs is resolved, GRAVITY will enable direct measurements of its mass by determining the dynamics of the broad line region. Measuring the velocity gradient allows to determine the broad line region (BLR) size, and  to test the scaling relation linking the luminosity of the AGN to the size of the BLR. GRAVITY can also be used to test nuclear star formation in AGN on scales 10 times smaller than what is possible nowadays with AO.
\item {\bf Young Stellar Objects} -- Spectro-interferometry in the K-band allows to probe the size and dynamics of the hydrogen Bracket gamma emission from the gas in the accretion disk across the whole initial mass function. GRAVITY will also study the morphology of YSO jets. Achieving 10 $\mu$as in one hour gives access to motions in the jets of T Tauri stars, moving at $\sim$ 150 km/s.
\item Furthermore, GRAVITY will allow studies of classical interferometric objects, such as binary stars, stellar surfaces, etc.
\end{itemize}
GRAVITY thus will carry out a number of fundamental experiments, as well as substantially increase the range and number of astronomical objects that can be studied with the VLTI thanks to its very high sensitivity and off-axis fringe tracking capability\footnote{A fringe tracker is the equivalent of an adaptive optics system for an interferometer. It measures and corrects the piston terms of the atmosphere between two telescope, which otherwise would blur the fringe signal.}.

\section{Overview of GRAVITY}

\begin{figure}
\centering
\includegraphics[width=0.8\textwidth]{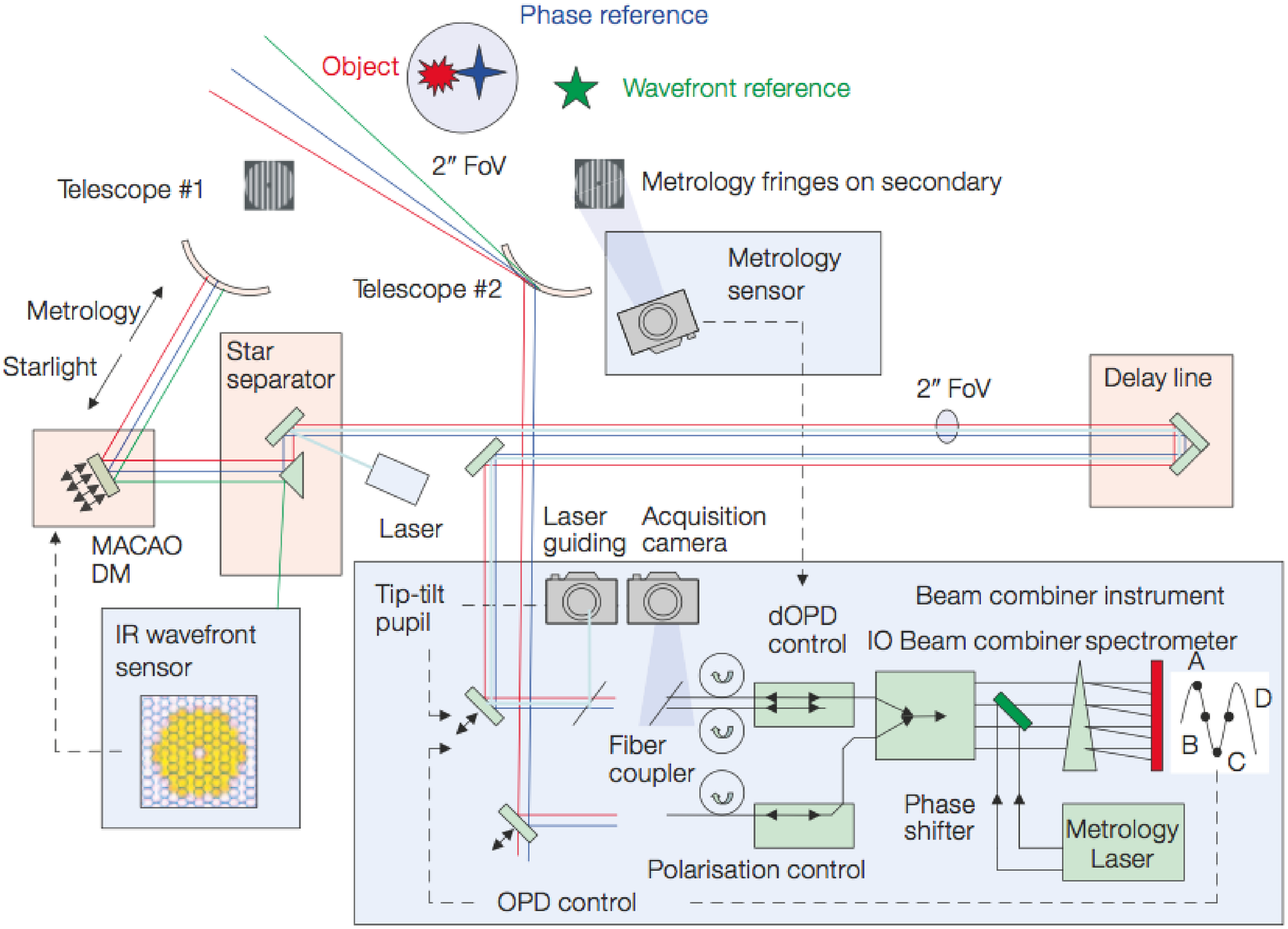}
\caption{Schematic overview of GRAVITY for one baseline.}
\label{fig:GRAVITY_overview}
\end{figure}

GRAVITY provides high precision narrow-angle astrometry and phase-referenced interferometric imaging in the astronomical K-band. It combines the light from four VLT telescopes (UTs or ATs), and measures the interferograms from 6 different baselines simultaneously. The instrument consists in several systems, pictured in Fig.~\ref{fig:GRAVITY_overview}: the beam-combiner instrument (consisting itself of several sub-systems), the laser metrology system \citep{gillessen_2012a} and the IR wavefront sensors \citep{kendrew_2012a}. A schematic overview of the whole instrument can be seen on Fig.~\ref{fig:GRAVITY_overview}.

\noindent {\bf AO} -- In order to reach its high sensitivity and stability requirement with the UTs, GRAVITY is assisted by four IR wavefront sensors mounted in the UTs Coud\'e room, picking one of the two beams provided by the star separators. They then control the MACAO deformable mirrors to provide a corrected PSF to the beam combiner instrument.

%
\noindent {\bf Beam combiner} -- The beam combiner is where the four beams of the VLTI are coherently combined and analysed. The system can work on one of the two beams behind the star separators. Any additional tip/tilt from the beam relay down to the VLTI laboratory is corrected by a dedicated laser-guiding system \citep{pfuhl_2012a}. Low frequency drifts of the field and pupil are corrected by the internal acquisition camera \citep{amorim_2012a}. These systems then provide a stable, corrected Point Spread Function (PSF) to the beam combiner instrument. The latter work on the 2-arcsecond VLTI field-of-view for UTs (4 arcseconds for ATs), containing both the reference star (used for fringe tracking) and the science object.

\noindent The light of the two objects from the four telescopes is coupled into single-mode optical fibres -- for modal filtering -- in the four Fibre Couplers \citep{pfuhl_2014a}. Light then goes through the fibre Control Unit (Sect.~\ref{part:FCU}) to compensate for the differential delay and to adjust the polarisation. The fibres feed two integrated optics beam combiners (Sect.~\ref{part:IOBC}) and the coherently combined light is dispersed in two spectrometers \citep{straubmeier_2014a}. A low resolution spectrometer provides phase- and group-delay tracking \citep{choquet_2014a} on the reference star, thus enabling long exposure on the science target. Three spectral resolutions (R = 20, 450 \& 4000) are implemented in the science spectrometer, and a Wollaston prism provides basic polarimetry. GRAVITY will measure the visibility of the reference star and the science object simultaneously for all spectral channels, and the differential phase between the two objects. This information will be used for interferometric imaging and astrometry using the differential phase and group delay on both linear polarisations. All functions of the GRAVITY beam-combiner instrument are implemented in a single cryostat for optimum stability, cleanliness, and thermal background suppression.

\noindent {\bf Metrology} -- The internal path lengths of the VLTI and GRAVITY are monitored using a dedicated laser metrology (Sect.~\ref{part:metrology}). The laser light is propagated backward, from the spectrometers through the beam combiners, fibre contol unit and fibre couplers, and covers the full VLTI beam up to the telescope spider above the primary mirror. A set of photo-diodes measures then the intensity of the temporally modulated metrology fringes, from which the path lengths are extracted.

\section{Fibre control unit}
\label{part:FCU}

The fibre control unit of GRAVITY was developed at the LESIA (Paris, France) by Perrin et al. It contains fluoride fibres that were produced by Le Verre Fluor\'e. The goal of the FCU is to maximize the interferometric efficiency of GRAVITY as well as to correct for the differential OPDs between the science and the fringe tracker stars, thanks to two subsystems: {\it 1)} The Fibered Differential Delay Lines; {\it 2)} The Fibered Polarisation Rotators.
GRAVITY counts eight of each sub-system: one per spectrometer and per telescope. A fibre solution was preferred for both systems because of modal filtering property and the higher throughput compared to a bulk optics solution. GRAVITY working in the K-band, fluoride glass fibres were selected. They are not polarization maintaining because the technology is not mature yet, and to avoid a systematic polarization splitting, required by the high birefringence (by design) that could lead to null the fringe contrast in some unfavorable cases.

\noindent {\bf Fibre Differential Delay Lines} -- Differential delay lines are necessary for GRAVITY to match the delays of its two fields of view, so as to measure fringes of maximum contrast (i.e.~maximum signal-to-noise ratio (SNR)). Moderate delay ranges are required as the maximum separation between sources is 4 arcsec with the Auxiliary Telescopes and 2 arcsec with Unit Telescopes. For the maximum envisioned baseline length of 200 m, this represents a maximum optical delay of 4 mm, which can be achieved by stretching a fibre. Fibered delay lines have already been used for lab experiments, as well as on the 1$^{st}$ generation 3-beam VLTI fringe tracker FINITO \citep{gai_2004}, though with a much more limited stroke (only few micrometers for internal fringe sampling purpose). The differential OPD stability between two FDDLs (one fixed, one tracking) must reach 50 nm, for not reducing the fringe contrast by more than 1\% in the science channel during long integration times of 1 minute.

\noindent {\bf Fibre Polarisation Rotators} --  The instrument must also be neutral with respect to polarizations for not destroying fringes on the detectors. Polarisation must therefore be controlled and aligned to the natural axes of the instrument. Optimization of fringe contrast using a fibered polarization rotator has been demonstrated with the FLUOR instrument \citep{coudeduforesto_1998a}, was also used on the instruments VINCI, and ‘OHANA, and was therefore selected for GRAVITY.

Laboratory characterisations showed that the fibre Control Unit has a throughput higher than 95\% over the full K-band (1.9-2.45 $\mu m$) on its eight channels, and is also regularly operated in the GRAVITY cryostat at a temperature of 240 K.

\begin{figure}
\centering
\includegraphics[width=0.98\textwidth, trim=0 17.5em 0 4em, clip=true]{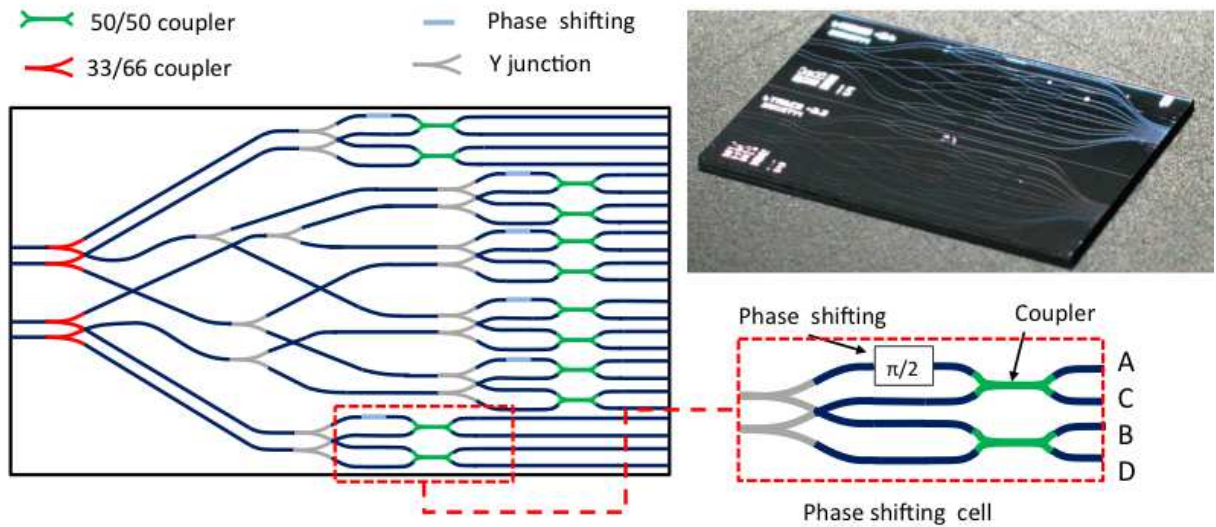}
\caption{Schematics and picture of the IOBC. The light from the four telescopes is injected from the left. The flux of each telescope is equally split between three baselines, thanks to a 33/66 coupler, plus a 50/50 Y-junction on the brightest arm. Telescopes are then coherently recombined by pair in a phase shift cell (lower part of the drawing). The cell contains an achromatic $\pi/2$ phase shift section, as well as symmetric couplers, resulting in an ABCD fringe sampling.}
\label{fig:IOBC}
\end{figure}

\section{Integrated optics beam combiners}
\label{part:IOBC}

The Integrated Optics Beam Combiners (IOBC) were developed jointly by IPAG (Grenoble, France) and LETI (Grenoble, France), and characterised at IPAG. They are part of the IOBC Assemblies, which include in addition the V-groove interface to the fibre Control Unit fibres (Sect.~\ref{part:FCU}) at their entrance, and the mount to the spectrometers. The IOBC Assembly was recently fully described in more details in \cite{jocou_2014a}.

GRAVITY uses two of these combiners, one for each spectrometer. They are operated under a pressure of $10^{-6}$ mbar, and at a temperature of 200 K. Each assembly combines four telescopes of the VLTI and form interferometric fringes of the six baselines in the K-band (1.95 -- 2.45 $\mu m$). Each possible baseline is combined in a pairwise fashion, like e.g.~PIONIER \citep{lebouquin_2011a}. The fringe on each baseline is then sampled in four points simultaneously, $\sim 90^\circ$ apart (the so-called ABCD fringe coding \citep{colavita_1999a}), leading to a total of 6$\times$4 = 24 outputs (see Fig.~\ref{fig:IOBC}). The outputs are then imaged on the detectors by the spectrometers, which provide the appropriate filters for a spectral and/or polarimetric analysis of the fringes.

The waveguides have a mode-field radius of 3.83 $\mu m$ at $\lambda =$ 2150 nm, and are therefore single-mode across the full K-band, down to 1.85 $\mu m$, allowing to transmit the metrology laser ($\lambda = 1908\:$nm) in a single-mode regime in the backward direction (from the spectrometers to the telescopes; Sect.~\ref{part:metrology}). To operate the metrology, the IOBC must in addition stands a laser power of about 1 W. The IOBC outputs are also anti-reflection coated for $\lambda = $1908 nm to minimize reflections towards the science detectors. The spectrometers contain an additional set of two line filters to block the rest of the metrology light to a level of $10^6$. From the characterisation in the lab, it was determined that the IOBCs have an average throughput of 53\% over the full K-band (including the fibre assembly) with a plateau reaching 70\% between 1.9 and 2.1 $\mu m$. The IOBC fringe contrast is always better than 94\% over the whole spectral band, leading to a high instrumental answer.

\begin{figure}[h]
\centering
\includegraphics[width=0.9\textwidth]{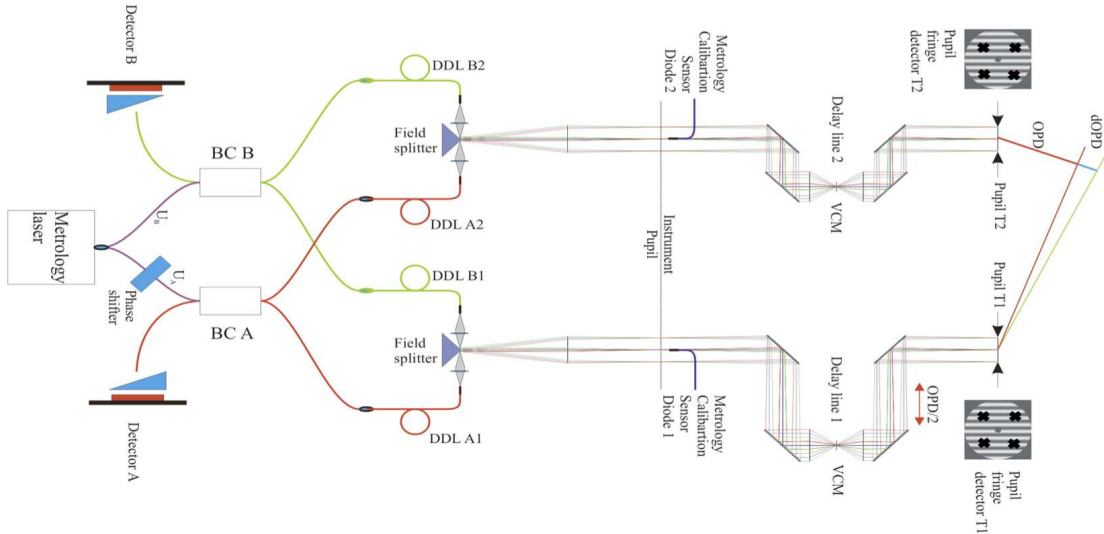}
\caption{Schematics of the GRAVITY metrology working principle.} \label{fig:metrology}
\end{figure}

\section{Metrology}
\label{part:metrology}

The metrology of GRAVITY is developed at the MPE. The astrometric requirement of GRAVITY of few 10 micro-arcseconds on sky implies that the metrology must measure the optical path difference between the two beam combiners of GRAVITY to a level of 5 nm. The design of the metrology is presented in details by \cite{gillessen_2012a}, and the last results of its integration and performance were reported in, e.g.,~\cite{kok_2014a}, \cite{lippa_2014a}, and \cite{blind_2014a}. The metrology laser ($\lambda = 1908$ nm) is sent backward into GRAVITY, from the spectrometers (FT and SC) up to the telescopes, so as to trace back the exact same optical paths that follows the stellar light (Fig.~\ref{fig:metrology}), i.e.~GRAVITY beam combiners, VLTI optical train, and telescope. Fringes are then formed in the primary mirror space and are detected with four photodiodes placed on the telescope spiders. The phase is finally measured by modulating an electro-optics phase shifter.

As reported during the conference, the metrology laser excites a fluorescence line in the 20 m of fluoride fibres of the beam combiner instrument, that were contaminated by Thulium and Holmium elements from a previous batch of doped fibres. The result is an illumination of the detectors on the blue part of the K-band (1950 to 2200 nm), reducing the instrument sensitivity by several orders of magnitude. During the last months, an upgrade of the metrology has been developed to solve this problem and is currently being implemented. A third {\it carrier} beam $I_c$ is now launched after the fibres, and interferes with the two others ($I_{FT}$ and $I_{SC}$), so that the fringe signal of each arm has now an amplitude $\sqrt{I_c \: I_{SC, FT}}$. Hence, increasing the carrier flux $I_c$, allows to reduce the spectrometer metrology fluxes and fluorescence in proportion, while preserving the SNR. The phase of each arm is now measured by modulating a different electro-optic phase shifter on each, and at different frequencies $f_{FT}$ and $f_{SC}$, in the kHz range. By applying appropriate sawtooth voltages, we generate two almost pure sine signals of the same two frequencies on each diode. Two dual channel lock-in amplifiers are locked on these frequencies, and detect the real and imaginary parts of the two fringe signals ($\cos(\varphi_{SC, FT})$ and $\sin(\varphi_{SC, FT})$). Each amplifier is then measuring the phase of each arm, $\varphi_{FT}$ and $\varphi_{SC}$, the difference of the two being the metrology phase. The narrower detection bandwidth ($\sim 100 Hz$) of this scheme also allows to increase the detection sensitivity, as well as to reject unwanted noise sources like vibrations.

These improvements allow to run the metrology with a laser power reduced by a factor $10^3$ - $10^4$ in the spectrometers. The fluorescence is now invisible from the FT spectrometer, and close to the sky background level on the SC, allowing to operate again the instrument in optimum conditions.

\section{Summary}

GRAVITY is an interferometric instrument making a massive use of fibre and integrated components to accomplish various optical functions (coherent beam transport and combination, polarisation control, phase shifting, etc.) with a high efficiency. It is now in its last integration and test phases. The beam combiner instrument and metrology system will be shipped to the VLT observatory in Chile in 2015, for a comissionning planned to start in October 2015.



\begin{thebibliography}{}
\expandafter\ifx\csname natexlab\endcsname\relax\def\natexlab#1{#1}\fi
\expandafter\ifx\csname url\endcsname\relax
  \def\url#1{\texttt{#1}}\fi
\expandafter\ifx\csname urlprefix\endcsname\relax\def\urlprefix{URL }\fi
\providecommand{\eprint}[2][]{\url{#2}}

\bibitem[Amorim et al.(2012)]{amorim_2012a} Amorim, A., Lima, J., 
Anugu, N., et al.\ 2012, in Proc. of SPIE, vol. 8445, 34A

\bibitem[{Blind et~al.(2014)}]{blind_2014a}
Blind N., Huber H., Eisenhauer F., et al., 2014, in Proc. of SPIE, vol. 9146, 24B

\bibitem[{{Choquet} et~al.(2014)}]{choquet_2014a}
{Choquet}, {\'E}., {Menu}, J., {Perrin}, et al., 2014, \aap, 569, A2.

\bibitem[{{Colavita} et~al.(1999)}]{colavita_1999a}
{Colavita}, M.~M., {Wallace}, J.~K., {Hines}, et al., 1999, \apj, 510, 505.

\bibitem[{{Coud{\'e} du Foresto} et~al.(1998)}]{coudeduforesto_1998a}
{Coud{\'e} du Foresto}, V., {Perrin}, G., {Ruilier}, et al., 1998, in Proc. of SPIE, vol. 3350, 856

\bibitem[{{Gai} et~al.(2004)}]{gai_2004}
{Gai}, M., {Menardi}, S., {Cesare}, et al., 2004, in Proc. of SPIE, vol. 5491, 528

\bibitem[{{Gillessen} et~al.(2012)}]{gillessen_2012a}
{Gillessen}, S., {Lippa}, M., {Eisenhauer}, et al., 2012, in Proc. of SPIE, vol. 8445, 1OG

\bibitem[{Jocou et~al.(2014)}]{jocou_2014a}
Jocou, L., Perraut, K., Moulin, T., et al., 2014, in Proc. of SPIE, vol. 9146, 1JJ

\bibitem[{Kendrew et~al.(2012)}]{kendrew_2012a}
Kendrew, S., Hippler, S., Brandner, W., et al., 2012, in Proc. of SPIE, vol. 8445, 7WK

\bibitem[{Kok et~al.(2014)}]{kok_2014a}
Kok, Y., Gillessen, S., Lacour, S., et al. 2014, in Proc. of SPIE, vol. 9146, 25K

\bibitem[{{Le Bouquin} et~al.(2011)}]{lebouquin_2011a}
{Le Bouquin}, J.-B., {Berger}, J.-P., {Lazareff}, et al., 2011, \aap, 535, A67

\bibitem[{Lippa et~al.(2014)}]{lippa_2014a}
Lippa, M., Blind, N., Gillessen, S., et al., 2014, in Proc. of SPIE, vol. 9146, 22L

\bibitem[{Pfuhl et~al.(2014)}]{pfuhl_2014a}
Pfuhl, O., Haug, M., Eisenhauer, F., et al., 2014, in Proc. of SPIE, vol. 9146, 23P

\bibitem[{{Pfuhl} et~al.(2012)}]{pfuhl_2012a}
{Pfuhl}, O., {Haug}, M., {Eisenhauer}, F., et al., 2012, in Proc. of SPIE, vol. 8445, 1UP

\bibitem[{Straubmeier et~al.(2014)}]{straubmeier_2014a}
Straubmeier, C., Yazici, S., Wiest, M., et al., 2014, in Proc. of SPIE, vol. 9146, 29S
  
\end{thebibliography}
\end{document}